\documentclass[%
 aip,
 amsmath,amssymb,
 reprint,%
]{revtex4-1}

\usepackage{graphicx}
\usepackage{dcolumn}
\usepackage{bm}

\usepackage[utf8]{inputenc}
\usepackage[T1]{fontenc}
\usepackage{mathptmx}
\usepackage{etoolbox}
\usepackage{xcolor}
\usepackage{comment}
\usepackage{CJK}

\def \aap  {A\&A}

\def \apj  {ApJ}
\def \apjs  {ApJS}
\def \apjl  {ApJL}

\def \mnras {MNRAS}

\def \physrep {Phys. Rept.}

\def \prl {PRL}

\makeatletter
\def\@email#1#2{%
 \endgroup
 \patchcmd{\titleblock@produce}
  {\frontmatter@RRAPformat}
  {\frontmatter@RRAPformat{\produce@RRAP{*#1\href{mailto:#2}{#2}}}\frontmatter@RRAPformat}
  {}{}
}%
\makeatother
\begin{document}
\begin{CJK*}{UTF8}{ipxg}

\newcommand{\Alfven}{Alfv\'{e}n\,}
\newcommand{\Alfvenic}{Alfv\'{e}nic\,}


\title[Ion Weibel Instability in the hybrid framework]{Ion Weibel Instability in the hybrid framework: the optimal resolution}
\author{Luca Orusa}
\affiliation{Department of Astronomy and Columbia Astrophysics Laboratory, Columbia University, 538 West 120th Street, New York, NY 10027, USA}
\affiliation{Department of Astrophysical Sciences, Princeton University, Princeton, NJ 08544, USA}
\author{Taiki Jikei (寺境太樹)}
\affiliation{Department of Astronomy and Columbia Astrophysics Laboratory, Columbia University, 538 West 120th Street, New York, NY 10027, USA}
\affiliation{Department of Earth and Planetary Science, The University of Tokyo, 7-3-1 Hongo, Tokyo 113-0033, Japan}

\email{luca.orusa@columbia.edu; t.jikei@columbia.edu}

\date{\today}

\begin{abstract}
The study of collisionless shocks and their role in cosmic-ray acceleration has gained increasing importance through both observations and simulations. Accurately modeling the shock transition region, where particle injection and energization occur, requires a proper description of the microinstabilities governing its structure. In high-Mach-number astrophysical shocks, such as those associated with supernova remnants, the ion Weibel instability is believed to provide the dominant dissipation mechanism. In this work, we investigate the ion Weibel instability driven by counterstreaming beams in the presence of an external perpendicular magnetic field, with beam velocities significantly exceeding the local \Alfven speed. We employ hybrid simulations, in which ions are treated kinetically while electrons are modeled as a charge-neutralizing fluid. Although hybrid models are widely employed to study collisionless shocks, the resolution requirements needed to accurately capture ion-scale instabilities remain poorly understood. We address this issue by developing a linear theory of the ion Weibel instability tailored to the massless electron assumption of hybrid models and validating it with one- and two-dimensional simulations over a wide range of \Alfvenic Mach numbers. We show that hybrid simulations can reliably reproduce the growth, saturation, and polarization of Weibel-generated magnetic fields in weakly magnetized regimes, provided that the relevant ion-scale modes are properly resolved. From the scaling of the dominant mode, we derive a minimum spatial resolution required as a function of \Alfvenic Mach number. We also demonstrate that excessive resolution introduces unphysical small-scale whistler modes inherent to the massless-electron approximation. We validate the analysis by comparing the results with full particle-in-cell simulations. Together, these results provide practical guidance for hybrid simulations of collisionless shocks and beam-driven plasma systems.
\end{abstract}

\maketitle

\section{Introduction}
\label{sec:intro}

Astrophysical shocks are widely regarded as the primary accelerators of cosmic rays, owing to the efficiency of well-established shock acceleration mechanisms \citep{Drury1983,Blandford1987}. 
In collisional systems, supersonic flows generate shocks through particle–particle collisions. However, most astrophysical environments consist of collisionless plasmas, where shocks must instead be mediated by plasma microinstabilities.

Among these, the Weibel instability\citep{Weibel1959,Fried1959} plays a key role, as it can generate strong magnetic fields even in the absence of a pre-existing background magnetic field. It is therefore considered a leading candidate for mediating high-Mach-number shocks, where the upstream kinetic energy density greatly exceeds the energy density stored in the ambient magnetic field.

The physics of Weibel-mediated shocks has been extensively investigated through theoretical studies and particle-in-cell (PIC) simulations, both in the relativistic regime \citep{Silva2003,Hededal2004,Frederiksen2004,Kato2005,Kato2007,Spitkovsky2008a,Spitkovsky2008b,Bret2013,vanthieghem+18} and in the non-relativistic regime relevant to supernova remnant (SNR) shocks \citep{Kato2008,kato+10,matsumoto+12,bohdan+21,Jikei2024a,vanthieghem+24}. Such shocks have also been explored experimentally in laser-driven plasma experiments \citep{Ross2012,Fox2013,Huntington2015,Fox2018,Fiuza2020}.


Despite its apparent simplicity as a beam-driven instability, the Weibel instability exhibits a complex dependence on multiple parameters: the beam species (ions or electrons), beam velocity, the temperatures of both the beam and background plasma, and the strength of the background magnetic field. Moreover, electron–ion shock simulations often adopt reduced mass ratios to alleviate computational costs. Owing to these factors, a unified understanding of the ion Weibel instability across the full parameter space remains incomplete, despite more than five decades of research.

Studying nonlinear shock physics without a firm understanding of the dominant instabilities is problematic, particularly since simulations and laboratory experiments often operate under conditions that differ significantly from realistic astrophysical environments. PIC simulations typically employ reduced dimensionality and mass ratios, while upstream conditions in laboratory laser experiments (e.g., temperature, magnetization, collisionality) differ from those in astrophysical settings \citep{Fox2018,Fiuza2020}. Assessing the impact of these approximations is therefore essential.

In this context, hybrid simulations, where ions are treated kinetically and electrons as a neutralizing fluid, provide a powerful tool to investigate the nonlinear physics of large-scale systems. By neglecting electron kinetic physics, these simulations focus exclusively on ion kinetic scales. This approach has enabled extensive simulation campaigns of 2D parallel shocks \citep{Caprioli2014a,Caprioli2014b,Caprioli2014c,Haggerty2020,caprioli+25} and 3D quasi-perpendicular shocks \citep{Orusa2023,Orusa2025,orusa+25}. Here, “parallel” and “perpendicular” refer to the angle between the direction of motion of the shock and the direction of the background field. However, relatively little attention has been devoted to determining the resolution requirements necessary for hybrid simulations to accurately capture the physics of the shock transition region, particularly in perpendicular configurations.

As shown in \citet{Orusa2025}, that focused on particle injection and acceleration in perpendicular shocks, numerical resolution has a substantial impact on the resulting particle spectra. In particular, the fraction of injected particles strongly depends on the downstream magnetic field topology, which itself is shaped by the instabilities governing the shock structure and how well they are resolved. A common but flawed assumption in previous studies has been that resolving turbulence on sub-$d_i$ scales is unnecessary, since the thermal ion gyroradius in the background magnetic field is $\gg \, d_i$. However, \citet{Orusa2025} demonstrated that under-resolving these smaller scales compromises the accurate modeling of key instabilities.

To date, an investigation of the ion Weibel instability within hybrid simulations remains absent from the literature. Addressing this gap is the primary goal of this work. In this paper, we investigate the ion Weibel instability in weakly perpendicular magnetized shocks using idealized hybrid simulations. We begin with a theoretical analysis of the linear growth of Weibel-generated magnetic fields within a hybrid framework that assumes massless electrons. We then validate this model using a suite of 1D hybrid simulations spanning a broad range of shock velocities and numerical resolutions. From these results, we derive criteria for the minimum resolution required to properly capture the relevant properties of the instability, criteria that should also inform the setup of shock simulations more generally.

We further demonstrate that increasing the resolution beyond a certain limit leads to the development of unphysical whistler modes, because of the assumption of massless electrons, which interfere with the correct representation of current filaments in the shock transition region. Based on this analysis, we provide a criterion for the maximum allowable resolution to avoid such unphysical effects and validate it through 2D hybrid simulations.


The paper is organized as follows.
The theory of the ion Weibel instability is described in Sec. \ref{sec:theory}.
Results of idealized 1D and 2D hybrid simulations are shown in Sec. \ref{sec:simulation}.
Finally, a summary and conclusions are given in Sec. \ref{sec:conclusion}.

\section{Theory} \label{sec:theory}
\subsection{Linear Theory}
We consider a symmetric beam of ions drifting in the $x$-direction, i.e., $n_0/2$ drifting with a bulk velocity of $+V_{\mathrm{sh}}\hat{\bm{e}}_x$, and $n_0/2$ drifting with $-V_{\mathrm{sh}}\hat{\bm{e}}_x$. 
The ambient magnetic field $B_0$ is set in the $z$-direction, perpendicular to the flows.
The wave vector is defined as $\bm{k}=(0, k\sin\theta, k\cos\theta)^{\top}$, which is motivated by the fact that the Weibel instability has a wave vector perpendicular to the flow direction. 
To derive the linear theory of the ion Weibel instability for the massless-electron model implemented in hybrid codes, we start from the standard treatment considering both ions and electrons, with masses $m_i$ and $m_e$, respectively. We then take the limit $m_i/m_e \rightarrow \infty$.
We normalize the plasma quantities in a \textit{hybrid-friendly} way: spatial scales are expressed in units of the ion skin depth, $d_i [= c/\omega_{\mathrm{p}}]$, where $c$ is the speed of light and $\omega_{\mathrm{p}} = \sqrt{4\pi n e^2/m_i}$ is the ion plasma frequency, with $m_i$, $e$, and $n$ denoting the ion mass, charge, and plasma number density, respectively. 
Time is measured in units of the inverse ion cyclotron frequency, $\omega_{\mathrm{c}}^{-1} [= mc/(eB_0)]$, where $B_0$ is the background magnetic field strength. 
We shall also define the \Alfvenic Mach-number $M_{\mathrm{A}}=V_{\mathrm{sh}}/V_{\mathrm{A}}$.
For this setup, the dispersion tensor derived from the linear theory of the ion Weibel instability reads:
\begin{widetext}
\begin{equation}
\bm{D} =
\begin{bmatrix}
\tilde{k}^2\cos^2\theta+1+\frac{m_{\mathrm{i}}}{m_{\mathrm{e}}}\frac{\tilde{\omega}^2}{\tilde{\omega}^2-(m_{\mathrm{i}}/m_{\mathrm{e}})^2} + \left(\frac{M_{\mathrm{A}}\tilde{k}}{\tilde{\omega}}\right)^2 & i\left(\frac{m_{\mathrm{i}}}{m_{\mathrm{e}}}\right)^2\frac{\tilde{\omega}}{\tilde{\omega}^2-(m_{\mathrm{i}}/m_{\mathrm{e}})^2} & -\tilde{k}^2\sin\theta\cos\theta \\
-i\left(\frac{m_{\mathrm{i}}}{m_{\mathrm{e}}}\right)^2\frac{\tilde{\omega}}{\tilde{\omega}^2-(m_{\mathrm{i}}/m_{\mathrm{e}})^2} & \tilde{k}^2\cos^2\theta+1+\frac{m_{\mathrm{i}}}{m_{\mathrm{e}}}\frac{\tilde{\omega}^2}{\tilde{\omega}^2-(m_{\mathrm{i}}/m_{\mathrm{e}})^2} & 0 \\
-\tilde{k}^2\sin\theta\cos\theta & 0 & \tilde{k}^2\sin^2\theta+\frac{m_{\mathrm{i}}}{m_{\mathrm{e}}}+1
\end{bmatrix},
\end{equation}
where $\tilde{\omega}=\omega/\omega_{\mathrm{c}}$ is the frequency in the unit of gyro-frequency, and $\tilde{k}=kd_\mathrm{i}$ is the wavenumber in the unit of reciprocal of the ion skin depth (see \citet{Jikei2024a} for details, note that, however, the \Alfvenic Mach number is defined using relative velocity of the beams in that article, which differs by a factor of 2).
By taking the $m_{\mathrm{i}}/m_{\mathrm{e}}\to\infty$ limit, we obtain the dispersion relation for the massless electron models:
\begin{equation} 
\lim_{m_{\mathrm{i}}/m_{\mathrm{e}}\to\infty}\bm{D}=\\
\begin{bmatrix}
\tilde{k}^2\cos^2\theta+1+\left(\frac{M_{\mathrm{A}}\tilde{k}}{\tilde{\omega}}\right)^2 & -i\tilde{\omega} & -\tilde{k}^2\sin\theta\cos\theta \\
i\tilde{\omega} & \tilde{k}^2\cos^2\theta+1 & 0 \\
-\tilde{k}^2\sin\theta\cos\theta & 0 & \infty
\end{bmatrix}
\end{equation}
\end{widetext}
Note that the $zz$-component is infinite, and it is not trivial how this should be treated.
Here, we focus on the wave vectors parallel to the ambient field $(\theta=0)$, which is the dominant direction for the Weibel instability.
In this case, the $zz$-component is decoupled from others, and we can consider the following dispersion equation,
\begin{equation}
\left[\tilde{k}^2+1+\left(\frac{M_{\mathrm{A}}\tilde{k}}{\tilde{\omega}}\right)^2\right]\left(\tilde{k}^2+1\right)-\tilde{\omega}^2=0.
\end{equation}

It is important to distinguish between the ion Weibel instability in unmagnetized plasmas and in weakly magnetized plasmas. 
Although the instability is driven by ions, the electron response plays a crucial role through what is commonly referred to as electron screening \cite{Achterberg2007,Ruyer2015}. 
The manner in which electrons respond to the induced electric fields determines both the characteristic transverse scale of the ion current filaments during the linear stage and the corresponding linear saturation level.
In fully unmagnetized plasmas, the electron response is controlled by electron inertia. 
Capturing this regime, therefore, requires a kinetic treatment of electrons and cannot be accurately modeled within standard hybrid simulations, where electrons are assumed to be massless.

In contrast, in weakly magnetized plasmas, where the characteristic timescale of the instability is shorter than the ion gyro-period but longer than the electron gyro-period, the electron dynamics are primarily governed by $\bm{E}\times\bm{B}$ drift and dynamo effects \cite{Jikei2024a}. 
In this regime, hybrid models can reproduce the electron response with reasonable accuracy, making them suitable for studying the ion Weibel instability under such conditions.

\subsection{Minimum Resolution} \label{subsec:minres}
\begin{figure}[t]
    \begin{center}
    \includegraphics[width=\linewidth]{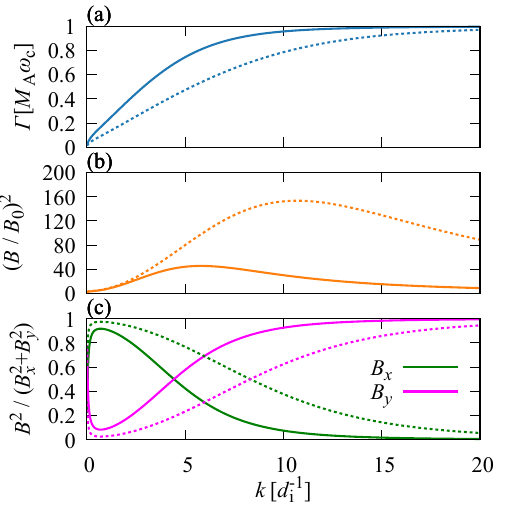}
    \caption{Linear theory for $M_{\mathrm{A}}=30$ (solid lines), and $M_{\mathrm{A}}=100$ (dotted lines).
    (a) Linear growth rate normalized by $M_{\mathrm{A}}\omega_{\mathrm{c}}$.
    (b) Estimated saturation level $(B/B_0)^2$, by Eq. (\ref{eq:saturation}).
    (c) Polarization $B^2/(B_x^2+B_y^2)$. 
    Green and magenta lines correspond to $B_x$ and $B_y$, respectively.} 
    \label{fig:linear_theory}
    \end{center}
\end{figure}

Figure~\ref{fig:linear_theory} presents the linear theory results for $M_{\mathrm{A}}=30$ (solid lines) and $M_{\mathrm{A}}=100$ (dotted lines). 
Panel (a) shows the linear growth rate $\Gamma$, normalized to $M_{\mathrm{A}}\omega_{\mathrm{c}}$.
The growth rate approaches $M_{\mathrm{A}}\omega_{\mathrm{c}}$, which is equivalent to $\omega_pV_{\mathrm{sh}}/c$.
As the \Alfvenic Mach number increases, the convergence to the maximum growth rate $\Gamma_{\mathrm{max}}=M_{\mathrm{A}}\omega_{\mathrm{c}}$ happens at a larger wavelength \cite{Jikei2024a}.
In the unmagnetized limit ($M_{\mathrm{A}}\to\infty$), the growth rate formally vanishes for all finite wavenumbers in the hybrid framework. 
In practice, however, electron inertia becomes important in this regime. 
When electron inertia is taken into account \cite{Jikei2024b}, the growth rate reaches $\omega_{\mathrm{pi}}$ at $k d_{\mathrm{i}} \sim \sqrt{m_{\mathrm{i}}/m_{\mathrm{e}}} \sim 43$.

\begin{figure}[t]
    \begin{center}
    \includegraphics[width=\linewidth]{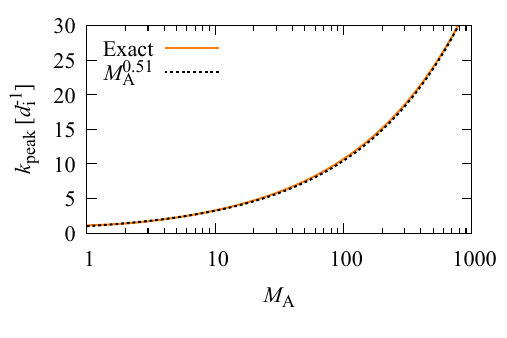}
    \caption{Dependence on $M_{\mathrm{A}}$ of the peak wavenumber. 
    The orange solid line shows the exact value, and the black dotted line is the best-fit power-law fitting $M_{\mathrm{A}}^{0.51}$.} 
    \label{fig:peak}
    \end{center}
\end{figure}

Figure~\ref{fig:linear_theory}(b) shows the estimated saturation of the Weibel modes due to the trapping mechanism \cite{Davidson1972}, which has been shown to provide a reasonable estimate for the saturation level at the end of the linear stage of the weakly magnetized Weibel instability \cite{Jikei2024a}.
The \textit{hybrid-friendly} form of this condition reads,
\begin{equation} \label{eq:saturation}
\left(\frac{B(k)}{B_0}\right)^2\sim 4M_{\mathrm{A}}^{2}\left(\frac{\Gamma}{M_{\mathrm{A}}\omega_{\mathrm{c}}}\right)^4(kd_{\mathrm{i}})^{-2}.
\end{equation}
For $M_{\mathrm{A}}=30$, the dominant wavenumber (solid line in Figure~\ref{fig:linear_theory}(b)), corresponding to the peak of the estimated saturation level, occurs at $k d_{\mathrm{i}} \sim 5.8$. 
This implies a peak wavelength of $\lambda_{\mathrm{peak}} = 2 \pi/k_{\rm peak} \sim 1.1 d_{\mathrm{i}}$. 
Empirically, resolving the nonlinear physics requires approximately $\sim 10$ cells per wavelength. Here, by ``empirically'' we refer to the commonly adopted numerical rule of thumb that a wave must be resolved with a sufficient number of cells in order to be accurately captured. We will test this hypothesis in Sec. \ref{sec:simulation}.
This translates into a minimum spatial resolution of $N=1/(10 \times 1.1) \sim 9$ cells per $d_{\mathrm{i}}$ to properly capture the Weibel instability in hybrid simulations at $M_{\mathrm{A}}=30$.
Comparing the solid line ($M_{\mathrm{A}}=30$) with the dotted line ($M_{\mathrm{A}}=100$) in Figure~\ref{fig:linear_theory}(b), we see that the peak wavenumber shifts to higher values as $M_{\mathrm{A}}$ increases. 
Since a larger peak $k$ corresponds to a shorter wavelength, higher spatial resolution is required in order to resolve this shorter wavelength.
At $M_{\mathrm{A}}=100$, for example, $\sim17$ cells per $d_{\mathrm{i}}$ is required.

Figure~\ref{fig:linear_theory}(c) shows the polarization $(B^2)/(B_x^2+B_y^2)$ during the linear stage.
The $B_x$ component is large at small wavenumbers, and the $B_y$ component becomes dominant at larger wavenumbers.
The transition happens around the peak wavenumber (see panel (b)).

Figure~\ref{fig:peak} shows the dependence of the peak wavenumber on $M_{\mathrm{A}}$. 
We find that $k_{\mathrm{peak}} \propto M_{\mathrm{A}}^{0.51}$ provides an excellent fit to the results obtained from the linear theory. 
This scaling leads to the following general expression for the minimum required spatial resolution:
\begingroup
\setlength{\abovedisplayskip}{8pt}
\setlength{\belowdisplayskip}{8pt}
\begin{equation}\label{eq:res_scaling}
N_{\rm min}=\frac{d_{\mathrm{i}}}{\Delta x_{\mathrm{min}}} = 9 \left(\frac{M_{\mathrm{A}}}{30}\right)^{0.51}
\end{equation}
\endgroup
or equivalently a resolution $\Delta x_{\mathrm{min}}=d_i/N_{\rm min}$ of:
\begingroup
\setlength{\abovedisplayskip}{8pt}
\setlength{\belowdisplayskip}{8pt}
\begin{equation}
\Delta x_{\mathrm{min}} = \frac{1}{9} \left(\frac{30}{M_{\mathrm{A}}}\right)^{0.51} d_{\mathrm{i}}
\end{equation}
\endgroup
In addition to the growth rate and the estimated saturation level, the linear theory also predicts the polarization of the magnetic fluctuations, defined as the ratio between $B_x^2(k)$ and $B_y^2(k)$ during the linear stage of the Weibel instability. 

\subsection{Maximum Resolution} \label{subsec:maxres}
\begin{figure}[t]
    \begin{center}
    \includegraphics[width=\linewidth]{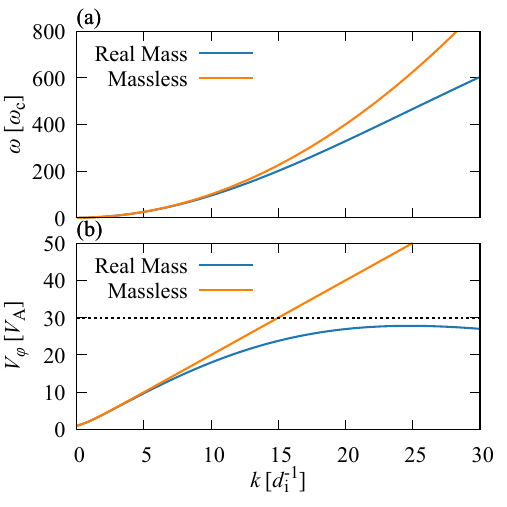}
    \caption{Characteristics of the whistler modes in hybrid codes. (a) Linear dispersion relation of the whistler mode. 
    The blue and the orange lines correspond to the dispersion relations with a realistic mass ratio and massless electrons, respectively. (b) Phase velocity with the same color code, overlayed with $V_{\phi}/V_{\mathrm{A}}=30$ in a black dotted line. Note that the phase velocity never exceeds 30 with the realistic mass ratio model.} 
    \label{fig:whistler}
    \end{center}
\end{figure}

Figure~\ref{fig:linear_theory}(c) presents the theoretical prediction, which will be compared with simulation results in Sec.~\ref{sec:simulation}.

One limitation of hybrid simulations is that increasing the resolution beyond what is physically required can introduce unphysical effects. 
In fact, since most hybrid models treat electrons as a massless, charge-neutralizing fluid, their behavior deviates from reality at scales approaching electron kinetic scales. 
A clear manifestation of this limitation is found in the properties of whistler modes.
Figure~\ref{fig:whistler} shows the dispersion relation of cold-plasma whistler modes.
Panel (a) presents $\omega$ as a function of $k$ for the realistic mass ratio $(m_{\mathrm{i}}/m_{\mathrm{e}}=1836)$ and for the massless-electron case $(m_{\mathrm{i}}/m_{\mathrm{e}}\to\infty)$. 
Panel (b) shows the corresponding phase velocity $\omega/k$ for both models. 
Significant deviations from the realistic case appear at $k d_{\mathrm{i}} \sim 20$. 
At resolutions of $\sim 30$ cells per $d_{\mathrm{i}}$, these wavelengths are resolved by $\sim 10$ cells, allowing unphysically large phase-velocity waves to develop in the simulations. 
In the massless-electron approximation, the electron cyclotron frequency is effectively infinite, leading to an overestimation of the phase velocity.
This may introduce spurious effects through artificial resonances, such as the modified two-stream instability \cite{Matsukiyo2003}, which would otherwise be stabilized for $M_{\mathrm{A}} > 30$ in realistic systems.

The theoretical considerations above, therefore, define an \textit{appropriate} resolution range for hybrid simulations employing the massless-electron approximation:
\begingroup
\setlength{\abovedisplayskip}{8pt}
\setlength{\belowdisplayskip}{8pt}
\begin{equation}\label{eq:limits}
9 \left(\frac{M_{\mathrm{A}}}{30}\right)^{0.51} 
< N < 30 
\end{equation}
\endgroup
or equivalently in resolution
\begingroup
\setlength{\abovedisplayskip}{8pt}
\setlength{\belowdisplayskip}{8pt}
\begin{equation}\label{eq:limits}
\frac{1}{9} \left(\frac{30}{M_{\mathrm{A}}}\right)^{0.51} d_{\mathrm{i}}> \Delta x > \frac{1}{30} d_{\mathrm{i}}.
\end{equation}
\endgroup

Note that the inequality on the right is independent of the \Alfvenic Mach number.
This range enables the study of Weibel-dominated systems while avoiding unphysical small-scale effects. 
These limits also allow us to identify a maximum $M_{\rm A} \sim 320$, beyond which the required resolution would introduce unphysical effects.
In other words, we can find an optimal resolution for $M_{\mathrm{A}}\ll300$.
This will be tested and validated with hybrid simulations in the following sections.
On the other hand, hybrid simulations may not be the right choice for $M_{\mathrm{A}}>300$.

\section{Simulation}
\label{sec:simulation}
All results presented in this work are obtained using the hybrid particle-in-cell code {\tt dHybridR} \cite{Haggerty2019}, which treats ions kinetically and electrons as a massless charge neutralizing fluid, in the non-relativistic regime \cite{Gargate2007}. 
Consistent with the setup described in Sec.~\ref{sec:theory}, we reproduce the theoretical configuration by simulating two symmetric, counterstreaming cold ion beams, each with density $n_0/2$ and velocity $\pm V_{\rm sh} \hat{e}_x$. The hybrid framework requires an explicit choice of electron equation of state; here, electrons are treated as an adiabatic fluid \cite{Caprioli2014a,Haggerty2020,Orusa2025} with index $\gamma = 5/3$.
We perform both 1D and 2D simulations. The 1D runs, used to analyze the linear theory of the ion-Weibel instability, resolve the $z$ direction, along which  

\begin{figure*}[t]
  \centering {
    \includegraphics[width=0.99\textwidth, clip=true,trim= 0 670 0 0]{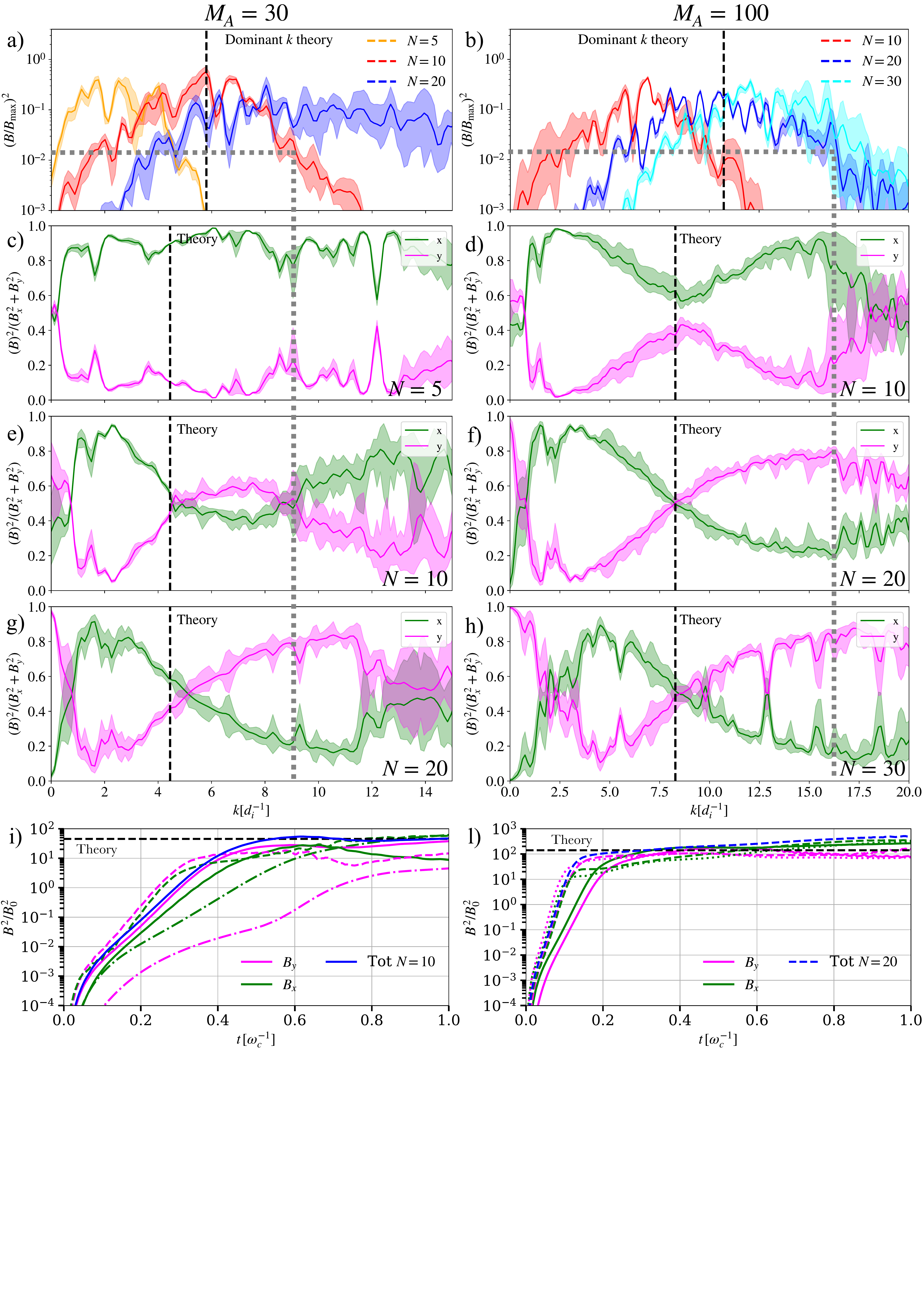}
  }
  \caption{
         Results from 1D hybrid simulations: panels a and b show the saturation level of $(B/B_{\max})^2$ for different $k$, with the line indicating the mean value and the shaded band showing the minimum and maximum values over different times. Panels c), d), e), f), g), h)  show the polarization $B^2/(B_x^2 + B_y^2)$ during the linear phase for simulations with 5 cells per $d_i$ (c), 10 cells per $d_i$ (d and e), 20 cells per $d_i$ (f and g) and 30 cells per $d_i$ (h). Panels i) and l) show the time evolution of the magnetic field generated by the Weibel instability for $N=5$ (dashed dotted line), $N=10$ (solid line), $N=20$ (dashed line) and $30$ (dotted line). In panels i) and l), we also show in blue the total generated magnetic energy, $B^2$, obtained for $N=10$ and $N=20$, respectively, together with the theoretical prediction (horizontal black dashed line) for the saturated value at the end of the linear phase obtained from Fig. \ref{fig:linear_theory}(b) and Eq. \ref{eq:saturation}. The left column corresponds to $M_{\rm A}=30$, and the right column to $M_{\rm A}=100$.
             }
             \label{fig:1D_simulation}
\end{figure*}
\clearpage
the background magnetic field is oriented, within a domain of $9\,d_i$. 
Simulations examining artificial whistler modes are instead carried out in 2D in the $x$–$z$ plane, within a $9 \times 9\, d_i^2$ domain. All simulations evolve the three components of particle momentum and electromagnetic fields. All simulations employ 8 particles per cell (ppc) and different number of cells $N$ per $d_i$. We also tested ppc$=32,64$, without any noticeable differences in the results.

\subsection{1D Simulation} \label{subsec:1D}

We begin by testing the theoretical prediction for the minimum resolution required to properly capture the ion Weibel instability using 1D simulations. We consider \Alfvenic Mach numbers $M_{\rm A} = 30$ and $M_{\rm A} = 100$.
We adopt three different resolutions: $N=5$, $N=10$, and $N=20$ cells per $d_i$ for $M_A=30$, and $N=10$, $N=20$, and $N=30$ cells per $d_i$ for $M_A=100$. For each of the two considered $M_A$ cases, these resolutions correspond respectively to: one case in which the instability is not properly captured, one that is marginally above the required theoretical threshold, and one with a resolution significantly larger than the minimum required value.

Two symmetric flows are initialized along the $\pm x$ direction, while the simulation resolves the $z$ direction, which also corresponds to the orientation of the background magnetic field. 

Results from the 1D hybrid simulations are shown in Figure~\ref{fig:1D_simulation}. 
Panels a) and b) display the power spectrum $(B/B_{\max})^2$ for different $k$ modes during the linear phase of the instability, where $B_{\max}$ is the maximum value of $B$ at each time. 
The linear phase is identified as the period during which the growth rate matches the theoretical prediction for the dominant $k$ modes discussed in Sec.~\ref{sec:theory}.

Panels c), d), e), f), g), h)  show the polarization $B^2/(B_x^2 + B_y^2)$ during the linear phase for simulations with 5 cells per $d_i$ (c), 10 cells per $d_i$ (d and e), 20 cells per $d_i$ (f and g) and 30 cells per $d_i$ (h). 
Vertical black lines indicate the theoretical predictions for the dominant $k$ mode in panels a) and b) and for the scale at which the polarization changes in panels c), d), e), f), g) and h).
Gray vertical lines mark the $k$ mode for which $(B/B_{\max})^2 = 10^{-2}$ for $N=10$ at $M_{\rm A}=30$ and for $N=20$ at $M_{\rm A}=100$, while the horizontal line indicates the threshold $(B/B_{\max})^2 = 10^{-2}$ used to identify these modes.

Panels i) and l) show the time evolution of the magnetic field generated by the Weibel instability for $N=5$ (dashed dotted line), $N=10$ (solid line), $N=20$ (dashed line) and $30$ (dotted line). In panels i) and l), we also show in blue the total generated magnetic energy, $B^2$, obtained for $N=10$ and $N=20$, respectively, together with the theoretical prediction (horizontal black dashed line) for the saturated value at the end of the linear phase obtained from Fig. \ref{fig:linear_theory}(b) and Eq. \ref{eq:saturation}.
The left column corresponds to $M_{\rm A}=30$, and the right column to $M_{\rm A}=100$.

As predicted by the theory, for $M_{\rm A}=30$ the dominant $k$ mode is properly resolved by $N=10$, and corresponds to the peak of the power spectrum obtained with this resolution.  
The $N=10$ simulations also reproduce the expected polarization behavior, including the transition to $B_y>B_x$ as a function of $k$ predicted by the linear theory shown in Figure \ref{fig:linear_theory}(c). 
At higher wavenumbers ($k > 10$), the polarization obtained with $N=10$ deviates from the theoretical prediction, with $B_y<B_x$; however, these modes contribute only $\sim 1\%$ of the total power, as shown in Figure~\ref{fig:1D_simulation}a), having a small impact on the overall behaviour of the instability and of the polarization. 
Overall, $N=10$ is sufficient to accurately capture the dominant mode. 
The higher resolution case, $N=20$, resolves smaller scales that are not necessary to capture the dominant mode at this Mach number. However, it produces more accurate results both in the power spectrum and in the polarization properties at high $k$. On the other hand, $N=5$ clearly under-resolves the dominant unstable mode and also fails to properly capture the polarization.
The magnetic field at the end of the linear phase reaches saturation levels of $\sim 42$, consistent with the prediction from the trapping mechanism (see also Figure~\ref{fig:linear_theory}(b)). While, in principle, $N=20$ would allow one to resolve even smaller scales, the goal of this paper is to provide a practical prescription for shock simulations that preserves optimal computational efficiency. In this context, we demonstrate that $N=10$ already provides sufficient resolution to capture the main characteristics of the instability. However, achieving higher precision at smaller scales requires correspondingly higher resolution.

For $M_{\rm A}=100$, on the other hand, $N=10$ is insufficient to capture the dominant $k$ mode, located at $k d_i = 10.71$, and a higher resolution is required. The power spectrum during the linear phase, shown in Figure~\ref{fig:1D_simulation}b, shows how $N=10$ underestimates the peak of the power spectrum with respect to linear theory, and that the predicted polarization is completely inconsistent with the linear-theory expectation.
On the other hand, $N=20$ properly captures the dominant $k$, and the peak of the power spectrum is consistent with the theoretical value, as is the polarization of the magnetic field, showing impressive agreement with the theory. Similarly, $N=30$ is able to correctly capture both the power spectrum and the polarization properties even at smaller scales. The magnetic field at the end of the linear phase reaches saturation levels of $\sim 140$, again consistent with the prediction from the trapping mechanism (see Figure also~\ref{fig:linear_theory}(b)). From Equation~\ref{eq:res_scaling}, the required resolution for $M_{\rm A} = 100$ must be at least $N=17$, confirming the results obtained in this section.
The simulations presented in this section demonstrate that the prediction obtained from Equation~\ref{eq:res_scaling} accurately identifies the minimum resolution required to capture the main characteristics of the instability for different $M_A$. However, achieving greater accuracy at smaller scales requires correspondingly higher resolution.

\subsection{2D Simulation} \label{subsec:2D}

\begin{figure}[t]
    \includegraphics[width=0.49\textwidth, clip=true,trim= 0 0 0 0]{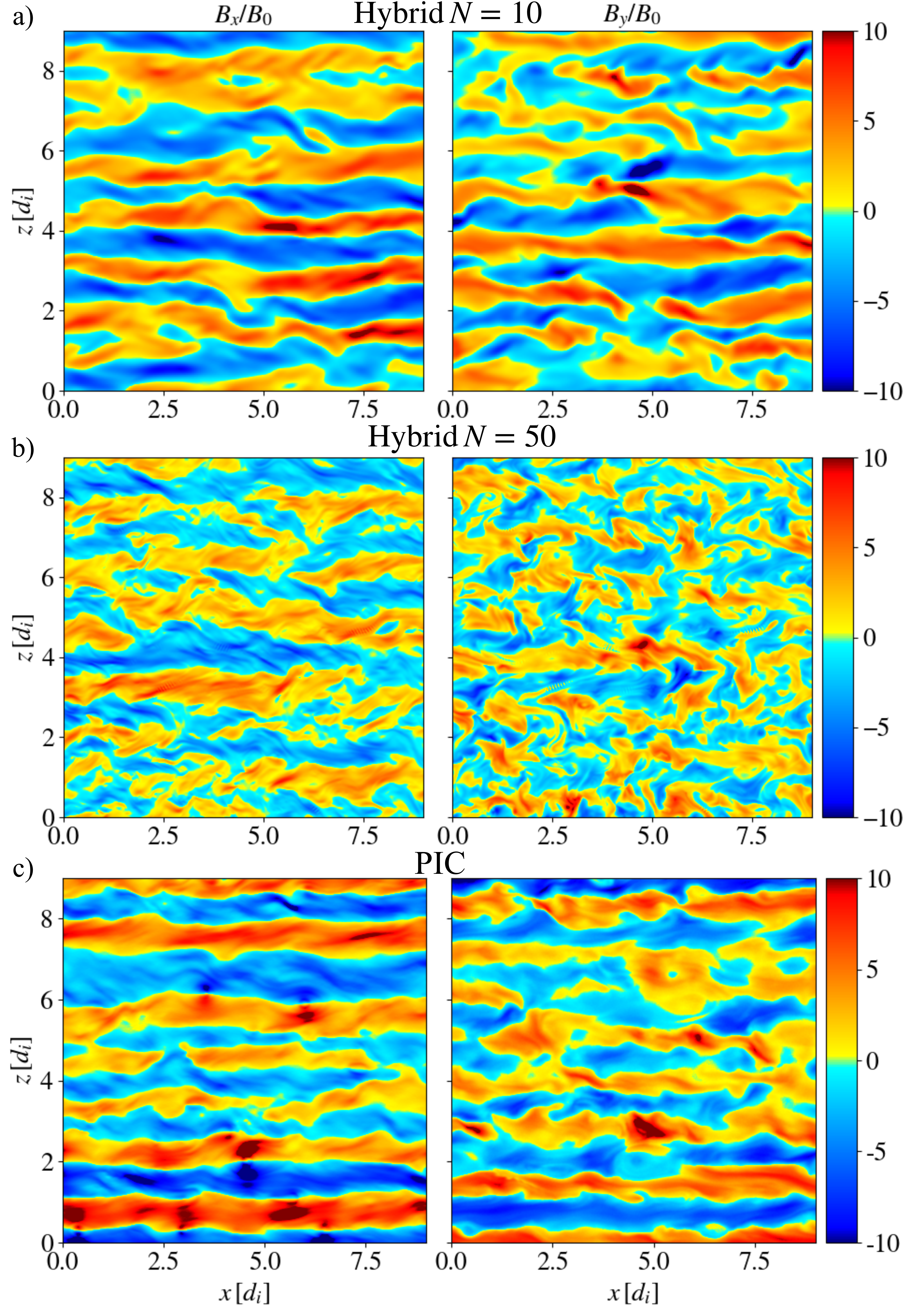}
  \caption{
         Results from 2D hybrid and PIC simulations for $M_A=30$ showing the $B_x$ and $B_y$ components: hybrid with $N=10$ (a), hybrid with $N=50$ (b), and full PIC with realistic mass ratio $m_{\rm i}/m_{\rm e} = 1836$ (c), all at saturation time $t = 0.6\,\omega_c^{-1}$.
             }
             \label{fig:2D_simulation}
\end{figure}

To test the maximum allowable resolution, we perform 2D simulations with $M_{\rm A} = 30$ in a $9 \times 9\,d_i^2$ box in the $x$--$z$ plane, with two counterstreaming flows moving along $x$, the background magnetic field oriented along $z$ and different resolutions. We then compare these results directly with a full PIC simulation performed under the same conditions, using a realistic mass ratio, $m_{\rm i}/m_{\rm e} = 1836$, and a resolution of 10 cells per $d_e$, which corresponds to 430 cells per $d_i$.
In Fig.~\ref{fig:2D_simulation} we show the $B_y$ and $B_x$ components generated by the Weibel instability at the same time, $t = 0.6 \, \omega_c^{-1}$, corresponding to the saturation of the linear phase, for hybrid with $N=10$ (a), $N=50$ (b) and full PIC (c).
These resolutions correspond, respectively, to one case in which the instability is marginally above the required theoretical threshold, and one with a resolution significantly larger than the maximum allowed value. In this way, we demonstrate that the limits identified in Equation~\ref{eq:limits} robustly distinguish the range of resolutions that properly capture the physical modes from those that produce unphysical behavior.

The $N=10$ hybrid simulation closely resembles the full PIC result, with filamentary structures of characteristic size $\sim d_i$. 
The PIC simulation naturally exhibits additional small-scale features due to its higher resolution and the inclusion of electron physics, but the ion-scale structures relevant for shock dynamics are well reproduced.

In contrast, for $N=50$, the magnetic field appears significantly more irregular, with oblique structures and without the well-defined filaments seen in both the full PIC and lower-resolution hybrid runs. The magnetic field fluctuations are also characterized by amplitudes $\Delta B/B_0 \ll 10$, in contrast to the full PIC and the $N=10$ cases. An additional simulation performed with $N=20$ produces a morphology very similar to both the $N=10$ hybrid and full PIC results and is therefore not shown. In contrast, for $N=5$ (also not shown, since the focus in this section is on the maximum resolution) the filaments develop on much larger spatial scales, the magnetic field becomes dominated by the $B_x$ component ($B_x>B_y$), and the simulation fails to reproduce the full PIC results.
A quantitative analysis of these waves is beyond the scope of this work, as such an investigation may be code-dependent. 
However, this comparison demonstrates that the theoretical prediction for the maximum useful resolution is robust, and that exceeding it can lead to the emergence of unphysical modes and results far from the PIC prediction.


\section{Conclusions}
\label{sec:conclusion}
In this work, we presented an investigation of the ion Weibel instability within the hybrid simulation framework and derived practical criteria for its accurate numerical representation.
Starting from a linear analysis appropriate for massless-electron models, we showed that hybrid simulations can correctly capture the ion-driven Weibel instability in weakly magnetized plasmas, provided that the relevant ion-scale modes are properly resolved. The dominant wavelength of the instability decreases with increasing \Alfvenic Mach number, leading to a minimum resolution requirement that scales approximately as $\propto M_{\rm A}^{0.51}$, where by minimum resolution we refer to the resolution for which the peak of the power spectrum corresponds to that predicted by linear theory, while also correctly reproducing the polarization properties at the dominant scales.
This establishes a physically motivated lower bound on the spatial resolution needed to reproduce the growth, polarization, and saturation of the instability.
At the same time, we demonstrated that increasing the resolution beyond $N=30$ cells per $d_i$ (see Eq. \ref{eq:limits}) can introduce artificial effects inherent to the hybrid approximation. In particular, resolving wavelengths approaching electron kinetic scales enables the growth of unphysical whistler modes with artificially high phase velocities.
The combination of these effects defines an optimal resolution range within which hybrid simulations faithfully reproduce the ion Weibel instability while avoiding nonphysical small-scale dynamics. This range also implies an upper limit on the Mach numbers ($\sim$ 320) that can be studied self-consistently within standard massless-electron hybrid models.

While treating electrons as a massless charge-neutralizing fluid has significant limitations in a high-Mach-number plasma environment, this may be relaxed by utilizing recently developed fluid models of plasmas.
Hybrid simulation models with finite electron mass effects have been developed \cite{Munoz2018, Jain2023}.
These models have a more accurate phase velocity of electron-scale waves, such as the whistler modes.
Small-scale dynamics would further be improved by incorporating kinetic effects into the electron fluid part \cite{Hammett1990, Hammett1992, Jikei2021, Jikei2022, Hunana2022, Hunana2025}.

Our results have direct implications for the modeling of collisionless shocks and other beam-driven plasma systems. As outlined in \citet{Orusa2025}, in perpendicular shocks, the magnetic topology of the transition region plays a key role in particle injection and acceleration, making it essential to accurately capture the ion-scale instabilities that shape it. 
The criteria derived here provide practical guidelines for designing hybrid simulations that are both computationally efficient and physically reliable. 
Future improvements in hybrid approaches, including more realistic electron closures, may extend the accessible parameter space and further enhance the fidelity of large-scale simulations of collisionless plasma dynamics.

\acknowledgments
We gratefully acknowledge T. Amano, D. Caprioli, L. Sironi, and A. Spitkovsky for helpful discussions. 
Simulations were performed on computational resources provided by the University of Chicago Research Computing Center.
L.O. acknowledges the support of the Multimessenger Plasma Physics Center (MPPC), NSF grants PHY2206607 and PHY2206609. T.J. is supported by grants from the Simons Foundation (MP-SCMPS-0000147) and NASA ATP 80NSSC24K1826.

\section*{Authors' Statements}
Conflict of Interest Statement: 
The authors have no conflicts to disclose.

Data Availability Statement:
The data that support the findings of this study are available from the corresponding author upon reasonable request.
\end{CJK*}
\bibliography{main}
\end{document}